\RequirePackage{ifpdf}

\newcommand{\commondocopts}{letterpaper,aps,prl,10pt,superscriptaddress,showpacs,floats,nofootinbib,twocolumn,lengthcheck}
\ifpdf
  \documentclass[\commondocopts,pdftex]{revtex4-1}
  \usepackage{cmap}
\else
  \documentclass[\commondocopts,ps2pdf]{revtex4-1}
\fi
\usepackage[latin1]{inputenc}
\usepackage[T1]{fontenc}
\usepackage{subfigure}
\usepackage{amsmath,amssymb,mathbbol}

\usepackage{graphicx,color}
\usepackage{hyperref}

\definecolor{rltred}{rgb}{0.75,0,0}
\definecolor{rltgreen}{rgb}{0,0.5,0}
\definecolor{rltblue}{rgb}{0,0,0.75}
\definecolor{rltdblue}{rgb}{0.2,0.2,0.65}
\definecolor{rltdred}{rgb}{0.65,0.2,0.2}
\definecolor{forestgreen}{rgb}{0.13,0.54,0.13}

\hypersetup{colorlinks,%
hypertexnames=true,%
pdfauthor={R. Pazourek, S. Nagele, J. Feist, J. Burgd\"orfer},%
pdftitle={Attosecond streaking of correlated two-electron transitions in helium},%
pdfview=FitH,%
pdfstartview=FitH,%
pdfpagemode=UseNone,%
bookmarksopen=true,%
bookmarksnumbered=true,%
pdfhighlight=/I,%
linkcolor=rltred,%
citecolor=rltgreen,%
linktocpage=true}

\begin{document}

\renewcommand{\equationautorefname}{Eq.}
\renewcommand{\figureautorefname}{Fig.}
\newcommand{\ui}{\mathrm{i}}
\newcommand{\ue}{\mathrm{e}}

\newcommand{\COMMENT}[1]{\textcolor{rltred}{\textbf{\textsc{#1}}}}
\newcommand{\etal}{\emph{et~al.}\ }
\newcommand{\ie}{i.e., }
\newcommand{\Schro}{Schr\"o\-din\-ger }
\newcommand{\eg}{e.g.\@ }
\newcommand{\cf}{cf.~}
\newcommand{\expval}[1]{\langle#1\rangle}
\newcommand{\abs}[1]{\left|#1\right|}

\newcommand{\bra}[1]{\langle#1|}
\newcommand{\ket}[1]{|#1\rangle}
\newcommand{\expt}[1]{\langle#1\rangle}

\newcommand{\refeq}[1]{\hyperref[#1]{\equationautorefname~(\ref*{#1})}}
\newcommand{\ddE}{\frac{\partial}{\partial E}}
\newcommand{\cvec}[1]{\mathbf{#1}}
\newcommand{\op}[1]{\mathrm{\hat{#1}}}
\newcommand{\vecop}[1]{\cvec{\hat{#1}}}
\newcommand{\eqcomma}{\,,}
\newcommand{\eqstop}{\,.}
\newcommand{\ed}{\,}
\newcommand{\hw}{\hbar\omega}

\newcommand{\ev}{\,\mathrm{eV}}
\newcommand{\eV}{\ev}
\newcommand{\au}{\,\mathrm{a.u.}}
\newcommand{\nm}{\,\mathrm{nm}}
\newcommand{\Wcm}{\,\mathrm{W}/\mathrm{cm}^2}
\newcommand{\as}{\,\mathrm{as}}
\newcommand{\fs}{\,\mathrm{fs}}
\newcommand{\He}{\mathrm{He}}
\newcommand{\Hep}{\He^+}
\newcommand{\tews}{t_{\scriptstyle\mathrm{EWS}}}
\newcommand{\tclc}{t_{\scriptstyle\mathrm{CLC}}}
\newcommand{\tislc}{t_{\scriptstyle\mathrm{ISLC}}}
\newcommand{\tfslc}{t_{\scriptstyle\mathrm{FSLC}}}
\newcommand{\tst}{t_{\scriptstyle\mathrm{S}}}
\newcommand{\tstcor}{t_{\scriptstyle\mathrm{S,c}}}
 
\title{Attosecond streaking of correlated two-electron transitions in helium}

\author{Renate~Pazourek}
\email{renate.pazourek@tuwien.ac.at}
\affiliation{Institute for Theoretical Physics, Vienna University of Technology, 1040 Vienna, Austria, EU}

\author{Stefan~Nagele}
\email{stefan.nagele@tuwien.ac.at}
\affiliation{Institute for Theoretical Physics, Vienna University of Technology, 1040 Vienna, Austria, EU}

\author{Johannes~Feist} 
\affiliation{ITAMP, Harvard-Smithsonian Center for Astrophysics, Cambridge, Massachusetts 02138, USA}

\author{Joachim~Burgd\"orfer}
\affiliation{Institute for Theoretical Physics, Vienna University of Technology, 1040 Vienna, Austria, EU}

\date{\today}

\begin{abstract}
We present fully ab initio simulations of attosecond streaking for ionization of helium accompanied by shake-up of the second electron. 
This process represents a prototypical case for strongly correlated electron dynamics on the attosecond timescale. 
We show that streaking spectroscopy can provide detailed information on the Eisenbud-Wigner-Smith time delay as well as on the infrared field dressing of both bound and continuum states. 
We find a novel contribution to the streaking delay that stems from the interplay of electron-electron and infrared-field interactions in the exit channel. We quantify all the contributions with attosecond precision and provide a benchmark for future experiments.

\end{abstract}
\pacs{32.80.Fb, 32.80.Rm, 42.50.Hz, 42.65.Re}

\maketitle

With the development of light-wave synthesis with attosecond precision time-resolved investigations of ultrafast electronic dynamics in atoms, molecules, and solids on the atomic time scale came into reach \cite{DreHenKie2001, KieGouUib2004, ItaQueYud2002}.
Attosecond streaking is one of the most fundamental processes in \emph{attosecond science} allowing for a mapping of time information onto the energy axis yielding a time resolution in the order of a few (tens) of attoseconds \cite{GouUibKie2004, DreHenKie2002, CavMueUph2007,SchFieKar2010}. 
It is a variant of a pump-probe setting with an ultrashort extreme ultraviolet (XUV) pulse serving as pump and a phase-controlled few-cycle infrared (IR) field as probe \cite{ItaQueYud2002}.
First proof-of-principle studies addressed the direct time-domain measurement of the life time of the Xe($4p^{-1}$) hole by Auger decay of $\simeq\!8$\,fs \cite{DreHenKie2002} and the time-resolved photoemission from a tungsten surface by an energetic XUV pulse of $\sim\!300\as$ duration \cite{CavMueUph2007} providing insight into the relative time delay of $\sim\!110\as$ \cite{CavMueUph2007} (later $\sim\!85\as$ \cite{CavKraErn2010}) between photoemission from core levels relative to the conduction band. 
Even shorter delays were more recently determined in the elementary photoelectric effect where a time shift of the $2p$ relative to the $2s$ electron in neon as small as $21\as$ has been obtained in an attosecond streaking experiment by Schultze \etal \cite{SchFieKar2010}.
These observations have triggered a flurry of theoretical investigations \cite{SchFieKar2010,KheIva2010,BagMad2010,NagPazFei2011,ZhaThu2010, YakGagKar2010,IvaSmi2011}.
While time delays of atomic photoionization on the one-electron (or mean-field) level as well as additional effective time shifts due to the dressing of the outgoing photoelectron ~\cite{ZhaThu2010,NagPazFei2011,IvaSmi2011} (often referred to as ``Coulomb-laser coupling'' (CLC), \cite{SmiMouPat2007}) could account for a delay of the order of $6\!-\!8\as$, even the approximate inclusion of correlation effects failed to reproduce the experimentally observed delay \cite{SchFieKar2010,KheIva2010}. 
However, multi-electron effects were only studied in stationary scattering approaches, \ie by analyzing the phase of the dipole transition matrix elements for photoionization in the absence of the streaking field. 
This \emph{group delay} or Eisenbud-Wigner-Smith (EWS) delay (see \eg \cite{deCNus2002}) is not directly what is measured by a streaking experiment \cite{NagPazFei2011}.
It was suggested that the probing IR field might be responsible for the larger delay observed in the experiment \cite{KheIva2010}.
There have been extensive time-dependent studies that simulated actual streaking experiments, but up to now only within the single-active electron approximation \cite{ZhaThu2010,NagPazFei2011}.
The role of electron correlations and its interplay with the laser field has remained an open problem. 
Treating both the dressing of the atomic dynamics by the IR streaking field and the dynamical electron correlation exactly is still out of reach for neon but is possible for helium.

In this letter we present streaking simulations for single ionization in helium including shake-up by solving the full two-electron time-dependent \Schro equation (TDSE) in the presence of the laser field without further approximation. 
In contrast to neon~\cite{SchFieKar2010}, where ionization from different atomic subshells was probed, shake-up ionization in helium relies strongly on electron-electron interactions. 
Here, the absorption of a single XUV photon leads to both emission of one electron and excitation of the other electron, which can only happen if the electrons share the photon energy and thus interact strongly. 
Studying ionization with and without shake-up may serve as benchmark for the interplay between mean-field, correlation, and field dressing effects. 

We show that for ionization without shake-up the streaking time shifts can be very well accounted for within the framework of Coulomb-laser coupling and the Eisenbud-Wigner-Smith delay. 
For ionization accompanied with excitation of the second electron correlation becomes important as field dressing effects modify the electron-electron interaction in the exit channel and give rise to an additional apparent time shift. 
We quantify all the contributions with attosecond precision and provide a benchmark for future experiments. 

\begin{figure}[tbh]
  \centering
  \includegraphics[width=\linewidth]{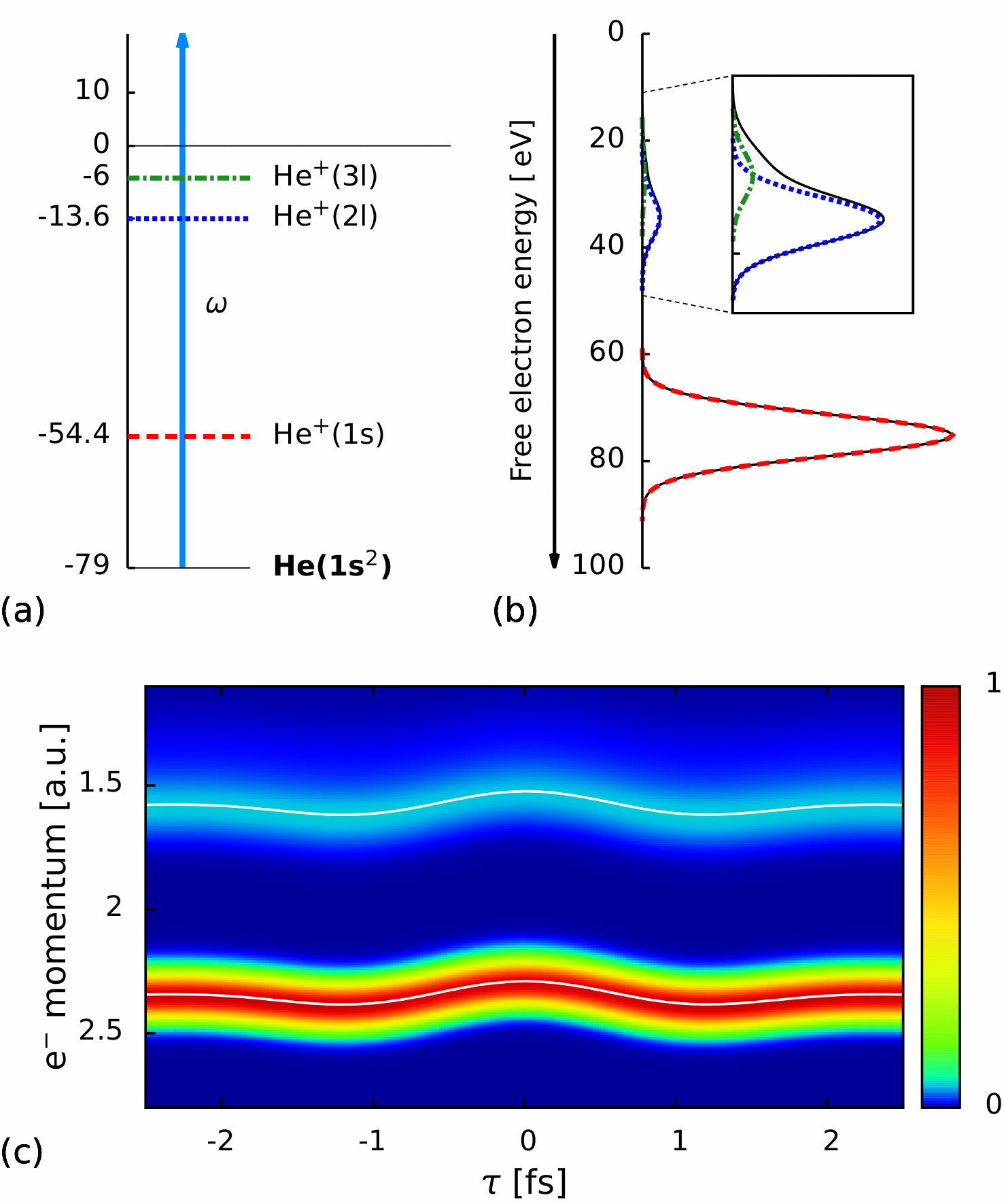}
  \caption{Single-ionization of He: (a) energy levels of $(nl, \varepsilon l')$ final states; (b) electron energy distribution $P^{SI}_{nl}(\varepsilon,\theta\!=\!0\pm10\textdegree)$ created by a $200\as$ (FWHM) XUV pulse with mean energy $\hw\!=\!100\ev$ along the polarization direction with an opening angle of 10\textdegree. (c) streaking spectrogram of (b) with a $3\fs$ streaking pulse with $\lambda\!=\!800\nm$ and intensity $I_{\scriptscriptstyle\mathrm{IR}}\!=\!4\cdot10^{11}\Wcm$, lower (high momentum) feature: shake-down, upper (low momentum): shake-up.
   } 
  \label{fig:schematics}
\end{figure}
In our computational approach (see \eg \cite{FeiNagPaz2008,SchFeiNag2011} for a detailed description) we solve the \Schro equation by the time-dependent close-coupling method, \cite{ColPin2002,LauBac2003,PinRobLoc2007}. 
For the radial discretization we use a finite-element discrete-variable representation (FEDVR) \cite{ResMcc2000,MccHorRes2001,SchCol2005}, and propagate in time using the short-iterative Lanczos (SIL) algorithm \cite{ParkLight86,SmyParTay1998} with automatic time-stepping and error control.
The laser fields are linearly polarized and treated in dipole approximation. The XUV pulse has a Gaussian envelope with a FWHM  duration of $200\as$ and an intensity of $I_{\scriptscriptstyle\mathrm{XUV}}\!=\!10^{12}\Wcm$ (for which multiphoton processes can be neglected), while the $800\nm$ IR field has a sine-squared envelope with a FWHM duration of $3\fs$ and an intensity $I_{\scriptscriptstyle\mathrm{IR}}\!=\!4\cdot10^{11}\Wcm$. 
For these laser parameters, numerically converged streaking simulations are obtained with a partial wave expansion with total angular momenta up to $L\!=\!8$ and one-electron angular momenta up to $l_\mathrm{<}\!=\!5$ and $l_\mathrm{>}\!=\!8$. 
We use an asymmetric radial box with an extension up to $960\au$ in one direction and $96\au$ for the other radial coordinate (which is enough for ionic bound states up to $n\!=\!8$ to be well represented). 
Each FEDVR element spans a length of $4.0\au$ and contains a DVR of order $11$. 
Atomic units are used throughout the paper.

\autoref{fig:schematics} shows the energy levels of helium and a corresponding single ionization spectrum including shake-up peaks (``correlation satellites''). 
Due to the large difference in the binding energy between the $\Hep(n\!=\!1)$ and $\Hep(n\!=\!2)$ states (\autoref{fig:schematics}a) of $1.5\au$ ($40.8\ev$) the two peaks are well separated and resolvable. 
For a typical XUV pulse ($\hw\!=\!100\ev$) with $200\as$ duration ($\approx\!9\ev$ spectral width), different shake-up channels (e.g. $n\!=\!2$ and $n\!=\!3$) are not resolved. However, numerically we have access to all channels separately. 
The electronic spectra for different delay times $\tau$ of the ionizing XUV pulse relative to the probing IR field build up a \emph{streaking spectrogram} (see \autoref{fig:schematics}c), where the spectra are shifted in momentum relative to the unperturbed asymptotic momentum $\vec{p}_0$. 
The streaking spectrograms are obtained by projecting the propagated wavefunction $\psi(\vec{r_1},\vec{r_2},t)$ onto the single continuum constructed as a symmetrized product state of a \emph{bound state} $\Phi_{n,l,m}(\cvec{r})$ of the $\He^+$ ion and a \emph{Coulomb wave} $\psi_{\cvec{k}}(\cvec{r})$ with charge $Z\!=\!1$. 

The absolute time shifts $\tst$ are extracted by a nonlinear least-squares fit of the modified final momentum $\vec{p}_f(\tau)$ of the different channels (taken as the first moment of the electron spectrum) 
\cite{NagPazFei2011}, 
\begin{equation}
\label{eq:streaking_simple}
\vec p_f(\tau) \approx \vec p_0 - \alpha \vec A_{\mathrm{IR}}(\tau+\tst) \, ,
\end{equation}
where $\alpha$ is a correction factor for the amplitude of the momentum shift induced by the streaking field. 
The resulting $\tst$ contains information on the Eisenbud-Wigner-Smith (EWS) time delay $\tews$ \cite{Eis1948,Wig1955, Smi1960} of the atomic photoionization process in the absence of the streaking field as well as on the dressing of atomic and ionic states in the IR field. 
On the one-electron level the streaking time shifts $\tst$ can be decomposed as \cite{NagPazFei2011}
\begin{equation}
\label{eq:streaking_delay1}
\tst = \tews + \tclc + \tislc \, .
\end{equation}

The EWS time delay $\tews$ is given by the energy derivative of the phase of the dipole transition element $\bra{\psi_f}z\ket{\phi_0}$, 
\begin {equation}
\label{eq:t_EWS}
 \tews(E)=\ddE \arg\left(\bra{\psi_f(E,\theta\!=\!0)}\hat z\ket{\psi_i}\right) \, ,
\end{equation}
which is evaluated along the laser polarization axis in forward direction, as is the streaking spectrogram.
The second term $\tclc$ is an apparent time shift due to the interaction of the outgoing electron with both the long-ranged Coulomb potential and the IR streaking field \cite{NagPazFei2011,ZhaThu2010,IvaSmi2011} which approximately scales with the final continuum energy $E$ as $\sim\!-E^{-3/2}$  \cite{Cla1979,NagPazFei2011}. 
We take $\tclc(E)$ from the \emph{reference} streaking shift $\tst^{\mathrm{H}(1s)}(E)$ of the pure $Z\!=\!1$ Coulomb potential at the asymptotic electron energy $E$, 
\begin{equation}
\label{eq:def_clc}
\tclc(E) = \tst^{\mathrm{H}(1s)}(E) - \frac{\partial}{\partial E}\sigma_1 \, ,
\end{equation}
where we have subtracted the EWS delay of the Coulomb phase $\sigma_1$ in the final $p$ continuum state for one-photon ionization of H(1s). 
The CLC time shifts are only weakly dependent on the streaking laser field parameters, in particular the wavelength of the IR field \cite{ZhaThu2010} and the duration of the XUV pulse. 

For strongly polarizable initial bound states another apparent time shift $\tislc$ was found due to initial-state laser coupling  \cite{NagPazFei2011,PazNagDob2011}. 
Energy shifts of the initial state due to the interaction with the laser field are at the moment of ionization transferred to the final energy, and thus appear as apparent time shifts. 

\begin{figure}[tb]
  \centering
  \includegraphics[width=\linewidth]{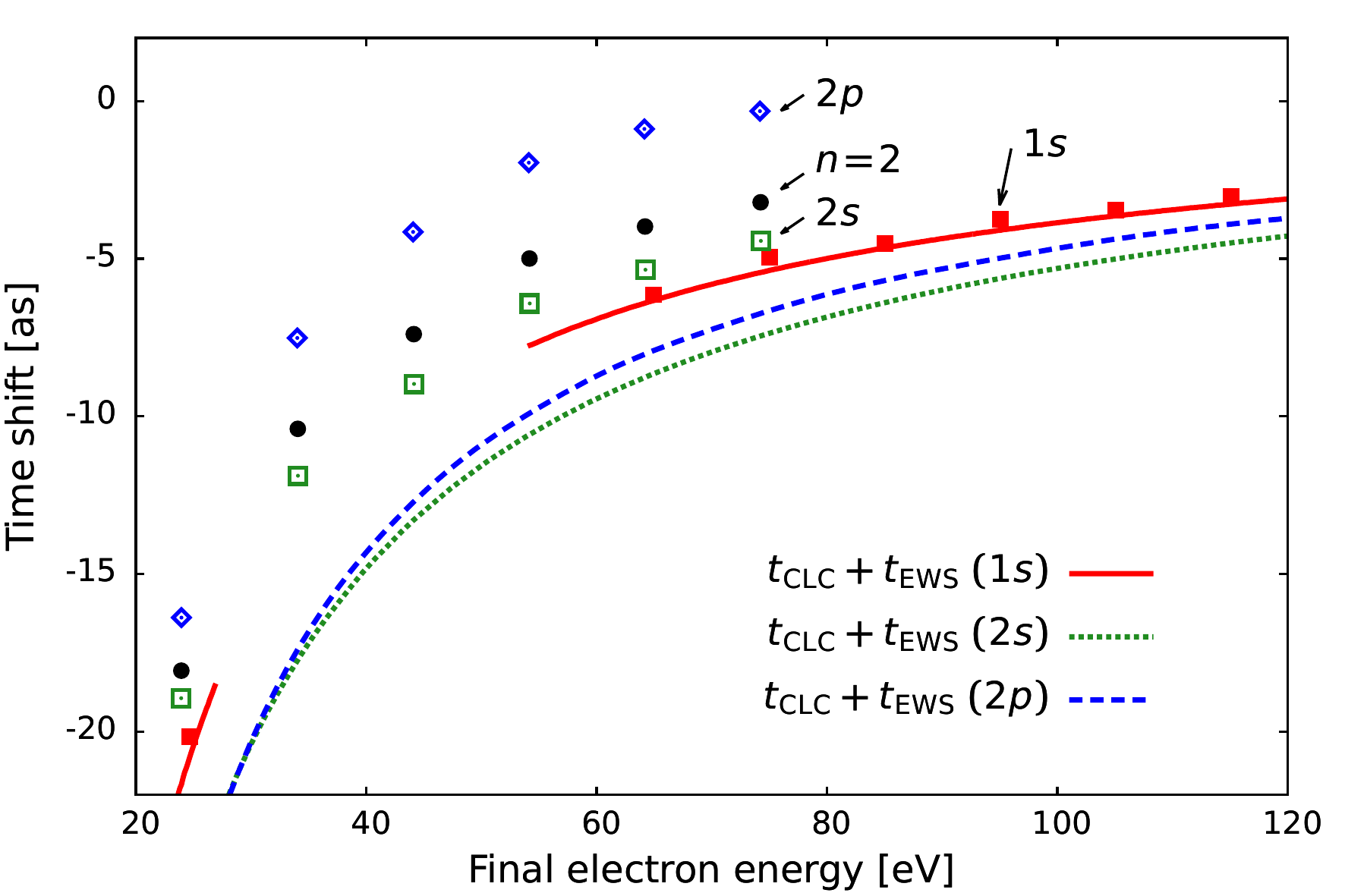}
  \caption{Temporal shifts $\tst$ (\small{\textcolor{red}{$\blacksquare$}}\,: 1s, \textcolor{forestgreen}{\boldmath$\square$}\,: 2s,  \textcolor{blue}{$\Diamond$}\,: 2p, $\bullet$\,: $n=2$) extracted from quantum mechanical streaking simulations and shifts predicted by \autoref{eq:streaking_delay1}, $\tclc+\tews$ (lines, with $\tislc\!=\!0$ for He($1s^2$)) for single ionization of helium into an opening angle of 10\textdegree with respect to the polarization axis with and without shake-up of the second electron as a function of the final electron energy. Time shifts that belong to the same XUV energy are thus shifted by $I_1^{(2)}-I_1^{(1)}\!=40.8\ev$. 
  Note that in the spectral region of resonances ($35\ev \leq E \leq 54.4\ev$ for an ionic $1s$ state), streaking time shifts are not well defined. } 
  \label{fig:timeshifts_n2}
\end{figure}

A novel scenario appears for true multi-electron processes beyond the single-active electron (SAE) or mean-field level. 
We consider the prototypical two-electron process, the photoionization of He with shake-up/down, where electron correlation is expected to play a significant role.
\autoref{fig:timeshifts_n2} shows the streaking time shifts $\tst$ for ionization with and without shake-up in comparison with the prediction $\tclc+\tews$ (\autoref{eq:streaking_delay1}). 
Initial state distortions $\tislc$ are negligible for the helium ground state He$(1s^2)$. 
For shake-down to the ground state in the ionic system $\Hep(1s)$, \ie direct ionization without shake-up, the experimentally accessible streaking shift $\tst$ agrees remarkably well with \autoref{eq:streaking_delay1}. 
To calculate exact dipole matrix elements for single ionization with the correct boundary conditions even above the double ionization threshold (\autoref{eq:t_EWS}), we use the extraction method of Palacios \etal\cite{PalMccRes2007} based on exterior complex scaling and apply it to the wave packet $\hat{z}\ket{\phi_0}$. 
For comparison we have also calculated the EWS time shift within the Hartree (mean-field) approximation where we take the ionic electron distribution to create an effective one-electron potential.
When the bound electron is left in the ionic ground state, also the mean-field values $\tews^{\mathrm{HF}}$ agree with the exact $\tews$ to within less than one attosecond (not shown).

\begin{figure}[tb]
  \centering
  \includegraphics[width=0.98\linewidth]{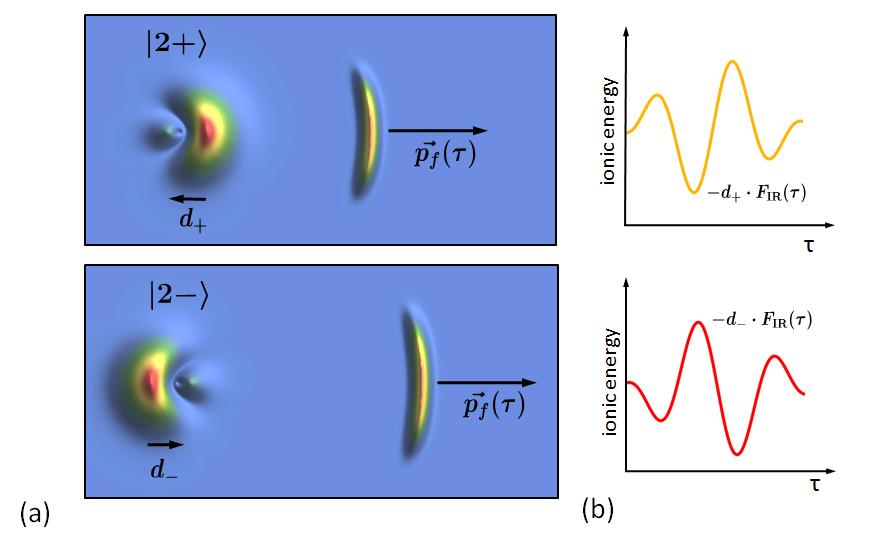}
  \includegraphics[width=\linewidth]{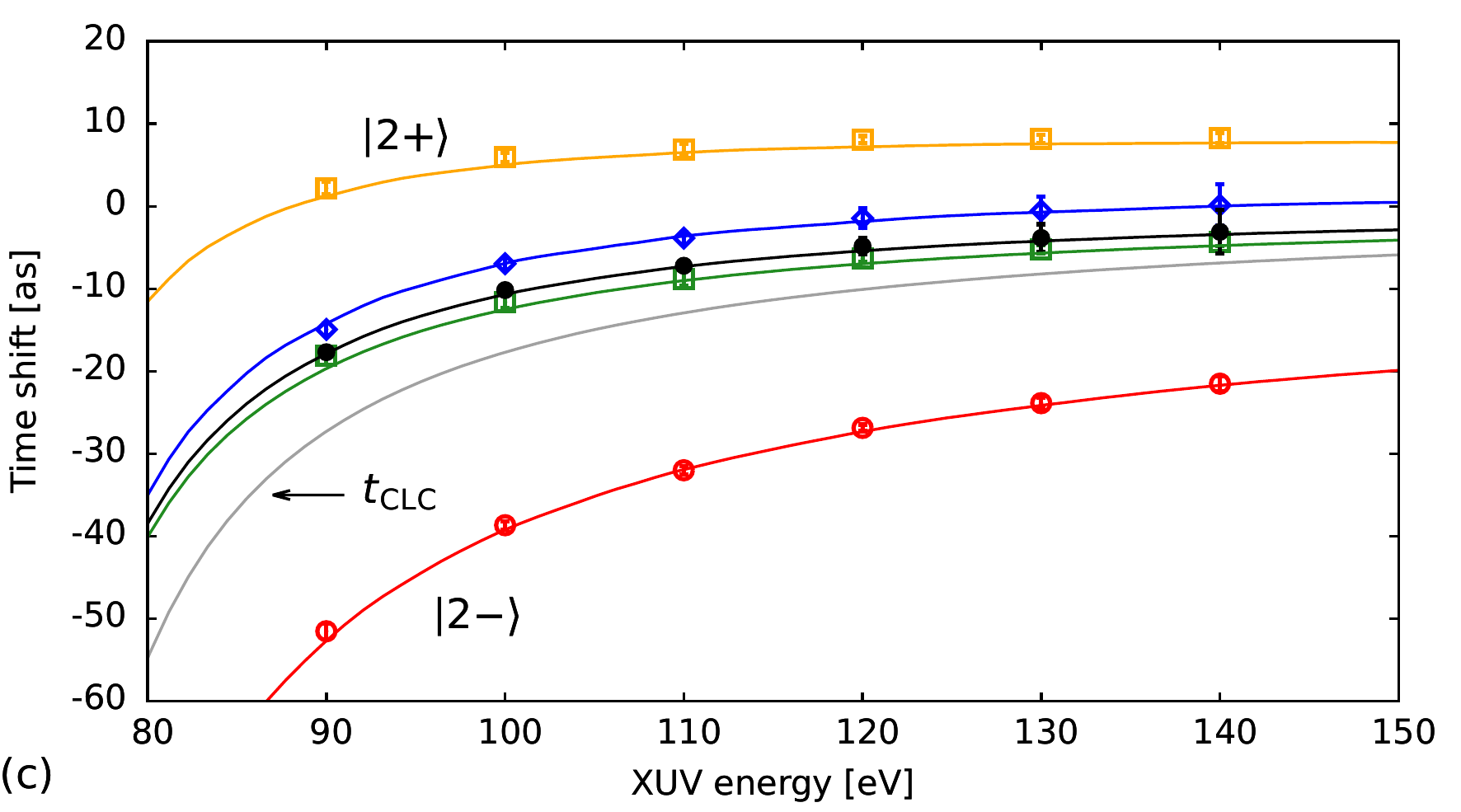}
  \caption{(a) Illustration of shake-up ionization with the ion remaining in a Stark state and (b) the ionic energy in the laser field which leads to an additional momentum shift of the outgoing electron. (c) Streaking time shifts $\tst$ (points) along the forward laser polarization direction for the two ionic parabolic states ($n\!=\!2, k\!=\!\pm1$) in comparison with $n\!=\!2$, $2s$ and $2p$ states (see \autoref{fig:timeshifts_n2}). In addition we show the complete prediction of $\tst$ according to \autoref{eq:streaking_delay2}, $\tclc+\tews+\tfslc$ (solid lines). 
  }
  \label{fig:he_stark}
\end{figure}

A surprisingly different picture emerges for shake-up to $n\!=\!2$ ($2s$ and $2p$) where $\tst$ and $\tclc+\tews$ strongly disagree (\autoref{fig:timeshifts_n2}). 
The streaking time shifts for shake-up predict a delay with respect to shake-down for all investigated photon energies which can not be accounted for by the corresponding EWS delays $\tews$. 
Obviously, the interplay between electron-electron interaction and the IR streaking field strongly influences the streaking delay of the outgoing wavepacket. 
The additional time delay can be viewed as the result of the coupling of the dipole moment of the shaken-up ionic state to the streaking field. 
Due to the entanglement of the system, the outgoing electron still contains information on the ionic state. 
For the almost-degenerate $n$ manifolds, this effect should be maximal for intershell eigenstates of the dipole operator, \ie Stark states $(nk)$ with $n\!=\!2$ and $k\!=\!\pm1$. 
Indeed, the resulting time delays (see \autoref{fig:he_stark}) are strongly enhanced $(k\!=\!1)$ and diminished $(k\!=\!-1)$ compared to ionic final states with well-defined angular momentum quantum number. 
There are two effects worth noting: First, the potential seen by the receding electron is modified by $V_D(\vec{r})=\vec{d}\!\cdot\!\vec{r}/r^3$, where $\vec{d}$ is the dipole moment of the Stark state.
However, this time shift is already included in the (field-free) EWS delay and thus gives no additional measurement-induced time shift. 
Second, creation of a shaken-up electron in a Stark state in the presence of the IR field entails an additional energy contribution $-\vec{d}\cdot\vec{F}_{\scriptstyle\mathrm{IR}}(t)$ of the ionic system (see \autoref{fig:he_stark}). 
This time-dependent energy shift of the \emph{bound} electronic final state of the residual ion becomes thus visible in the energy and momentum distribution of the streaked \emph{ionized} electron. 
Since this energy shift is proportional to $\vec{F}_{\scriptstyle\mathrm{IR}}(t)$ it appears as an additional time shift $\tfslc^{(2\pm)}=\mathrm{atan}\left(-\omega_{\scriptstyle\mathrm{IR}} d_z^{\pm}/k_z\right)/\omega_{\scriptstyle\mathrm{IR}}$. 
This process can be viewed as a two-electron generalization of the streaking shift for one-electron states with permanent dipole moment first discussed in \cite{BagMad2010}.
For final ionic Stark states the streaking delay $\tst^{(2\pm)}$ is indeed given to remarkable accuracy by the generalization of \autoref{eq:streaking_delay1},
\begin{equation}
\label{eq:streaking_delay2}
\tst = \tews + \tclc + \tislc + \tfslc \, 
\end{equation}
(with $\tislc=0$ for the He($1s^2$) initial state). 
This streaking-field induced two-electron effect causes also the additional time shifts for final ionic states with well defined angular momentum, $s$ and $p$, \ie states with a negligible dipole moment due to inversion symmetry. 
If we expand the final ionic states into Stark states, $\ket{2s/p}=1/\sqrt{2}\left(\ket{2+}\pm\ket{2-}\right)$ we can find an effective dipole moment for the final two-electron states, 
\begin{equation}\label{eq:mueff}
d_{z,\mathrm{eff}}^{l}=(d^{+} \abs{c_+}^2+d^{-} \abs{c_-}^2)/(2\abs{c_l}^2) \, , 
\end{equation}
where $c_\alpha=\bra{2\alpha,E\theta_0}z\ket{\psi_i}$. 
Note that the non-zero dipole moment in \autoref{eq:mueff} is consistent with the inversion symmetry of the ionic \emph{bound} state as this symmetry is broken by the selection of the emission direction of the streaked \emph{continuum} electron. 
Even if the final state of the remaining ion is not detected, \ie the time shift for the total $n=2$ shell is measured, an averaged effect in the order of $5\as$ is visible (see \autoref{fig:he_stark}). 
Thus probing the momentum spectrum (or energy shift, respectively) of the ionized electron contains information on the dynamics of the remaining ion. 

We have thus identified an additional time shift $\tfslc$ resulting from the back action of the excited bound state onto the continuum wave packet in the presence of the IR streaking field. 
Similar to the corresponding initial-state distortions, this additional time delay is not (or only weakly) dependent on the intensity of the IR field $I_{\scriptscriptstyle\mathrm{IR}}$ (within the range of useful intensities $10^{10}\Wcm\leq I_{\scriptscriptstyle\mathrm{IR}}\leq10^{12}\Wcm$), \cf \cite{NagPazFei2011}. 
Thus, for photoionization of two-electron systems with shake-up, the total streaking time shift (\autoref{eq:streaking_delay2}) contains an additional contribution due to simultaneous excitation of the second electron. 
Attosecond streaking phases can therefore reveal information on dynamical polarization in two-electron transitions. 
This information on the time-resolved bound-state excitation complements alternative techniques of transient x-ray absorption spectroscopy \cite{GouLohWir2010} and transient bound state excitation by shaped two-color laser fields \cite{XieRoiKar2011}. 

\begin{figure}[tb]
  \centering
  \includegraphics[width=\linewidth]{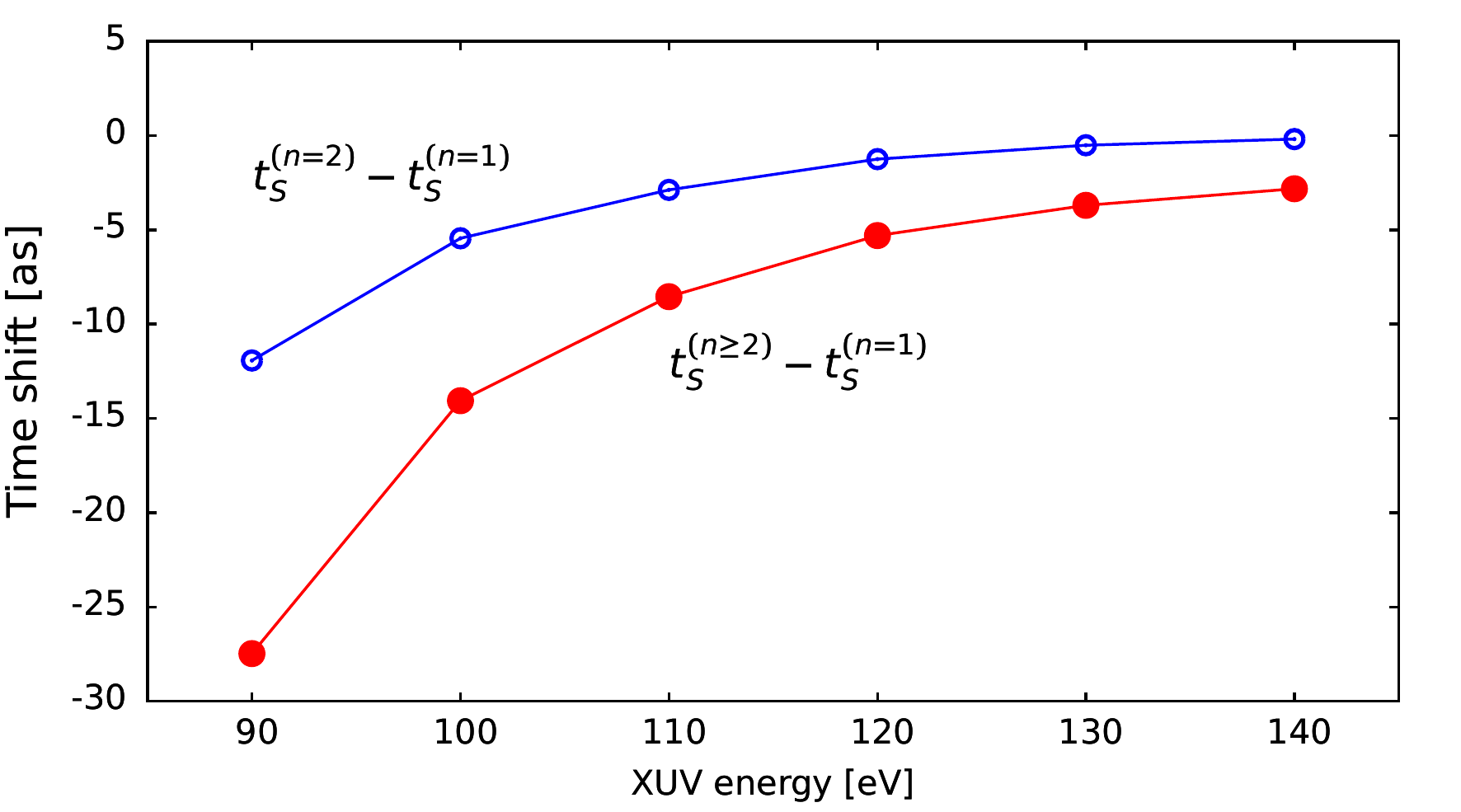}
  \caption{Relative total temporal shifts $\tst^{(n>1)}(\omega)-\tst^{(n\!=\!1)}(\omega)$ between ionization with and without shake-up of the second electron extracted from quantum mechanical streaking simulations for helium for different XUV energies, for shake-up into $n\!=\!2$ (blue line, open points) and sum of all shake-up states ($n\!\geq\!2$, red line, solid points), lines to guide the eye.}
  \label{fig:timeshifts_xuvdif}
\end{figure}

We conclude by discussing possible experimental observations. 
While the streaking spectrogram for the final ground state, $\Hep(1s)$, is well separated from that of two-electron excitation-ionization (\autoref{fig:schematics}), unambiguous observation of the latter requires the separation of the $n\!=\!2$ from higher shells with $n\geq\!3$. 
This would require an XUV pulse with a Fourier-limited width corresponding to a duration of $T_{XUV}\geq\!500\as$ (we have checked that our numerical results change within less than $1\as$ as compared to the $200\as$ used above). 
Alternatively, measuring Ly$_{\alpha}$ photons resulting from the radiative decay of the excited $n\!=\!2$ states in coincidence would allow to separate excitations of the $n\!=\!2$ shell. 
Moreover, excitations of the $2s$ and $2p$ states could be separated by prompt vs.~delayed Ly$_{\alpha}$ coincidences in a weak DC electric field. 
A simple proof of principle experiment would be to monitor the large delay of shake-up into the $n\!=\!2$ subshells, or an ensemble of excited manifolds ($n\!\geq\!2$), with respect to streaking without shake-up ($n\!=\!1$) as a function of the XUV energy (\autoref{fig:timeshifts_xuvdif}). 
This would give rise to time advances up to $12\as$ when the $n\!=\!2$ signal can be separated from higher shake-up states and up to $27\as$ when the total electron spectrum for $n\!>\!1$ is analyzed. 
The large difference in the delay for $n\!=\!2$ and $n\!\geq\!2$ shows how sensitive the extracted streaking time shifts are to small contributions of different shake-up states in the streaking spectrum (see inset \autoref{fig:schematics}). 
The strong photon energy dependence of the relative delay between $n\!=\!1$ and $n\!>\!1$ is due to the variation of $\tclc$ with the final-state energy $E$.
The ensemble of excited manifolds with $n\!\geq\!2$ would correspond to the streaking delay between the two peaks of the total ionization spectrum (see \autoref{fig:schematics}b, black line) which should be relatively straight-forward and could serve as benchmark for the precision of the experimental streaking techniques.

In summary, we have shown that for photoionization of helium streaking time shifts provide detailed information on the Eisenbud-Wigner-Smith time delay as long as Coulomb-laser coupling as well as laser-induced state distortion effects are accounted for. 
For two-electron excitation-ionization the interplay of electron-electron and IR-field interaction in the exit channel leads to additional and novel contributions to the time shift of the outgoing wavepacket. 
We show that dynamical correlation effects can play a significant role for attosecond streaking experiments.
Our theoretical ab-initio results can serve as an accurate benchmark for experimental attosecond streaking setups.

This work was supported by the FWF-Austria, Grant No.\ P21141-N16, P23359-N16, and SFB 041 (VICOM), the COST Action CM0702 (CUSPFEL), and in part by the National Science Foundation through TeraGrid/XSEDE resources provided by NICS and TACC under Grant TG-PHY090031. The computational results presented have also been achieved in part using the Vienna Scientific Cluster (VSC). RP acknowledges support by the TU Vienna Doctoral Program Functional Matter and JF acknowledges support by the NSF through a grant to ITAMP.

\end{document}